\documentclass[aps,twocolumn,english,superscriptaddress,citeautoscript,showkeys,preprintnumbers,amsmath,amssymb,floatfix,footinbib]{revtex4-2}
\usepackage{hyperref}
\hypersetup{colorlinks=true,linkcolor=red,citecolor=blue}
\usepackage{amsmath}
\usepackage{graphicx}
\usepackage{dcolumn}
\usepackage{xcolor}
\usepackage{float}
\usepackage{soul}
\usepackage{adjustbox}
\usepackage{lipsum}
\usepackage{blindtext}

\usepackage{lipsum}

\mathchardef\mhyphen="2D

\begin{document}

\title{Quantum Griffiths singularity in the stoichiometric heavy-fermion system CeRh$_4$Al$_{15}$}
\author{Rajesh Tripathi}
\email{rajeshtripathi@jncasr.ac.in}
\affiliation{ISIS Facility, STFC, Rutherford Appleton Laboratory, Chilton, Oxon OX11 0QX, United Kingdom}
\affiliation{Jawaharlal Nehru Centre for Advanced Scientific Research, Jakkur, Bangalore 560064, India}

\author{D. T. Adroja}
\email{devashibhai.adroja@stfc.ac.uk}
\affiliation{ISIS Facility, STFC, Rutherford Appleton Laboratory, Chilton, Oxon OX11 0QX, United Kingdom}
\affiliation{Highly Correlated Matter Research Group, Physics Department, University of Johannesburg, Auckland Park 2006, South Africa}

\author{Y. Muro}
\email{ymuro@pu-toyama.ac.jp}
\affiliation{Center for Liberal Arts and Sciences, Faculty of Engineering, Toyama Prefectural University, Imizu
939-0398, Japan}

\author{Shivani Sharma}
\affiliation{ISIS Facility, STFC, Rutherford Appleton Laboratory, Chilton, Oxon OX11 0QX, United Kingdom}

\author{P. K. Biswas}
\affiliation{ISIS Facility, STFC, Rutherford Appleton Laboratory, Chilton, Oxon OX11 0QX, United Kingdom}

\author{T. Namiki}
\affiliation{Graduate School of Science and Engineering, University of Toyama, Toyama 930-8555, Japan}

\author{T. Kuwai}
\affiliation{Graduate School of Science and Engineering, University of Toyama, Toyama 930-8555, Japan}

\author{T. Hiroto}
\affiliation{Research Network and Facility Services Division, National Institute for Materials Science, Tsukuba 305-0047, Japan}

\author{A. M. Strydom}
\affiliation{Highly Correlated Matter Research Group, Physics Department, University of Johannesburg, Auckland Park 2006, South Africa}
\affiliation{Max Planck Institute CPfS, 40 Nöthnitzerstr., D-01187 Dresden, Germany}

\author{ A. Sundaresan}
\affiliation{Jawaharlal Nehru Centre for Advanced Scientific Research, Jakkur, Bangalore 560064, India}

\author{S. Langridge}
\affiliation{ISIS Facility, STFC, Rutherford Appleton Laboratory, Chilton, Oxon OX11 0QX, United Kingdom}

\date{\today}

\begin{abstract}
We present a detailed investigation of the stoichiometric CeRh$_4$Al$_{15}$ single crystal compound using the temperature dependence of the heat capacity [$C_{\text{P}}$($T$)], electrical resistivity [$\rho$($T$)], magnetic susceptibility [$\chi$($T$)], and magnetization [$M$($H$)] measurements for a magnetic field ($H$) applied in the basal plane and along the $c$-axis. The low temperature power-law behavior of $C$/$T$ $\propto$ $\chi$ $\propto$ $T^{-1+\alpha}$, the isotherm magnetization, $M \sim H^{\alpha}$ with the exponent $\alpha$ = 0.45 - 0.55, and the $T$-linear resistivity $\Delta \rho$ $\sim$ $T^\epsilon$ with $\epsilon \sim$ 1 are found to be consistent with the formation of quantum Griffiths singularities in the non-Fermi-liquid (NFL) regime. We further investigated the spin dynamics of a polycrystalline sample of CeRh$_4$Al$_{15}$, using zero-field (ZF) and longitudinal-field (LF) muon spin relaxation ($\mu$SR) measurements. ZF-$\mu$SR measurements do not reveal any sign of long-range magnetic ordering down to 70~mK. The electronic relaxation rate ($\lambda$) below 0.5~K increases rapidly and shows a thermal activation-like characteristic [$T$log($\lambda$)$\sim$ $T$] over the entire measured temperature range between 70~mK to 4~K, indicating the presence of low energy spin fluctuations in CeRh$_4$Al$_{15}$. LF-$\mu$SR measurements show a time-field ($t/H^{\eta}$) scaling of the $\mu$SR asymmetry indicating a quantum critical behavior of this compound. Furthermore, inelastic neutron scattering study on the polycrystalline sample reveals two crystal field excitations near 19 and 33~meV. These features collectively provide strong evidence of NFL behavior in CeRh$_4$Al$_{15}$ due to the formation of Griffiths phase close to a $T$ $\rightarrow$ 0~K quantum critical point.


\end{abstract}

\maketitle
\section{INTRODUCTION}
\label{Intro}

When a long-range ordering of magnetic moments is suppressed by competing interactions, novel ground states of matter may emerge near the magnetic instability close to zero temperature~\cite{RevModPhys.79.1015}. For instance, the magnetic metals near $T_{\text{N}}\rightarrow$ 0, also called a quantum critical point (QCP), exhibit a breakdown of Fermi liquid (FL) behavior, termed as non-Fermi liquid (NFL) that challenges our current understanding of strongly correlated electron systems. The existence of such a NFL behavior has been proposed in many U, Yb, or Ce-based $f$-electron metals where the NFL behavior is found when a long-range magnetic ordering suppresses to zero temperature by a non-thermal control parameter such as tuning the chemical composition~\cite{PhysRevB.63.134411,Maple1995,PhysRevB.98.165136}, applying a pressure~\cite{mathur1998} or a magnetic field~\cite{doi:10.1126/science.1063539}. 

A growing number of compounds are displaying NFL behavior by chemical substitution where magnetic ordering disappears inhomogeneously in the vicinity of the QCP~\cite{RevModPhys.73.797,doi:10.1080/001075199181602,Gegenwart2008,doi:10.1126/science.1191195,doi:10.1146/annurev-conmatphys-062910-140546}. For example, the NFL behavior has been observed in CeCu$_{6-x}$Au$_x$~\cite{PhysRevLett.79.159}, or UCu$_{5-x}$Pd$_x$~\cite{PhysRevLett.75.2023}, in which the NFL behavior is interpreted as a distribution of Kondo temperatures~\cite{Miranda_1996}. Furthermore, a number of chemically substituted systems have been described within the context of Griffiths singularities near a magnetic instability, where spin-spin interactions freeze the localized $f$ moments, having low value of Kondo temperature ($T_{\text{K}}$), into clusters with a wide distribution of sizes; the larger clusters dominate the susceptibility and lead to divergent behavior of the thermodynamic quantities as the temperature is lowered~\cite{PhysRevLett.81.3531,PhysRevB.62.14975}. This model has been used to explain the NFL behavior of many $f$-electron chemical substituted systems such as Th$_{1-x}$U$_x$Pd$_2$Al$_3$, Y$_{1-x}$U$_x$Pd$_3$, or UCu$_{5-x}$Pd$_x$~\cite{PhysRevLett.81.5620}; Ce(Ru$_{1-x}$Rh$_x$)$_2$Si$_2$~\cite{PhysRevB.70.144415}, CePtSi$_{1-x}$Ge$_x$~\cite{PhysRevB.70.024401}.

On the other hand, a clean system near a QCP shows NFL behavior either by suppressing magnetic ordering by applying hydrostatic pressure, such as CeIn$_3$~\cite{mathur1998magnetically} or CePd$_2$Si$_2$~\cite{mathur1998magnetically,PhysRevB.61.8679} or even at ambient pressure (very few systems), such as CeNi$_{2}$Ge$_{2}$~\cite{PhysRevLett.82.1293}, CeRhBi~\cite{doi:10.7566/JPSJ.87.064708}, UBe$_{13}$~\cite{PhysRevLett.73.3018}, $\beta$-YbAlB$_{4}$~\cite{doi:10.1126/science.1197531}. In these cases, a zero-temperature paramagnetic to antiferromagnetic quantum phase transition (QPT) has been suggested as the possible source of NFL behavior~\cite{PhysRevB.48.7183}. 

Recent work reporting the discovery of the stoichiometric Kondo lattice heavy fermion system CeRh$_4$Al$_{15}$ suggests the possibility that this system may exhibit quantum criticality without tuning~\cite{NESTERENKO20191061}. The low temperature electronic specific heat shows a huge value of the Sommerfield coefficient $\gamma$ = 2~J/K$^2$mol with no magnetic long-range order ($\theta_P=-160 $ K) down to 50~mK~\cite{NESTERENKO20191061}. Hence, CeRh$_4$Al$_{15}$ is regarded as a nonmagnetic, heavy-fermion metal close to an antiferromagnetic (AFM) instability with strong AFM fluctuations. Considering that this system is stoichiometric and devoid of chemical disorder, its low-temperature behavior differs from that of clean systems. These observations, therefore, motivated us in-depth study of single and polycrystalline samples of CeRh$_4$Al$_{15}$ using a wide range of bulk and microscopic techniques at ambient pressure.

In order to understand the physics of NFL behavior in stoichiometric CeRh$_4$Al$_{15}$, we have carefully characterized the single crystal sample using heat capacity [$C_{\text{P}}$($T$)], electrical resistivity [$\rho$($T$)], magnetic susceptibility [$\chi$($T$)], and magnetization [$M$($H$)] measurements. We further investigated the spin dynamics of CeRh$_4$Al$_{15}$ polycrystalline sample using muon spin relaxation ($\mu$SR) and inelastic neutron scattering (INS) measurements. Both the bulk and microscopic results indeed suggest the NFL behavior driven by the quantum Griffiths phase (QGP) scenario over a wide temperature and magnetic field range. We show that the QGP in CeRh$_{4}$Al$_{15}$ emerges from the "Al atoms with partial site occupancy", which locally modify the exchange interactions and lead to the formation of magnetic clusters. This results in NFL behavior as proposed by Castro Neto $et~ al$.~\cite{PhysRevLett.81.3531,PhysRevLett.81.5620}. Furthermore, the INS shows crystal field excitations at 19 and 33~meV, indicating the localised nature of the Ce ions.

\begin{figure}
		\includegraphics[width=8.5cm, keepaspectratio]{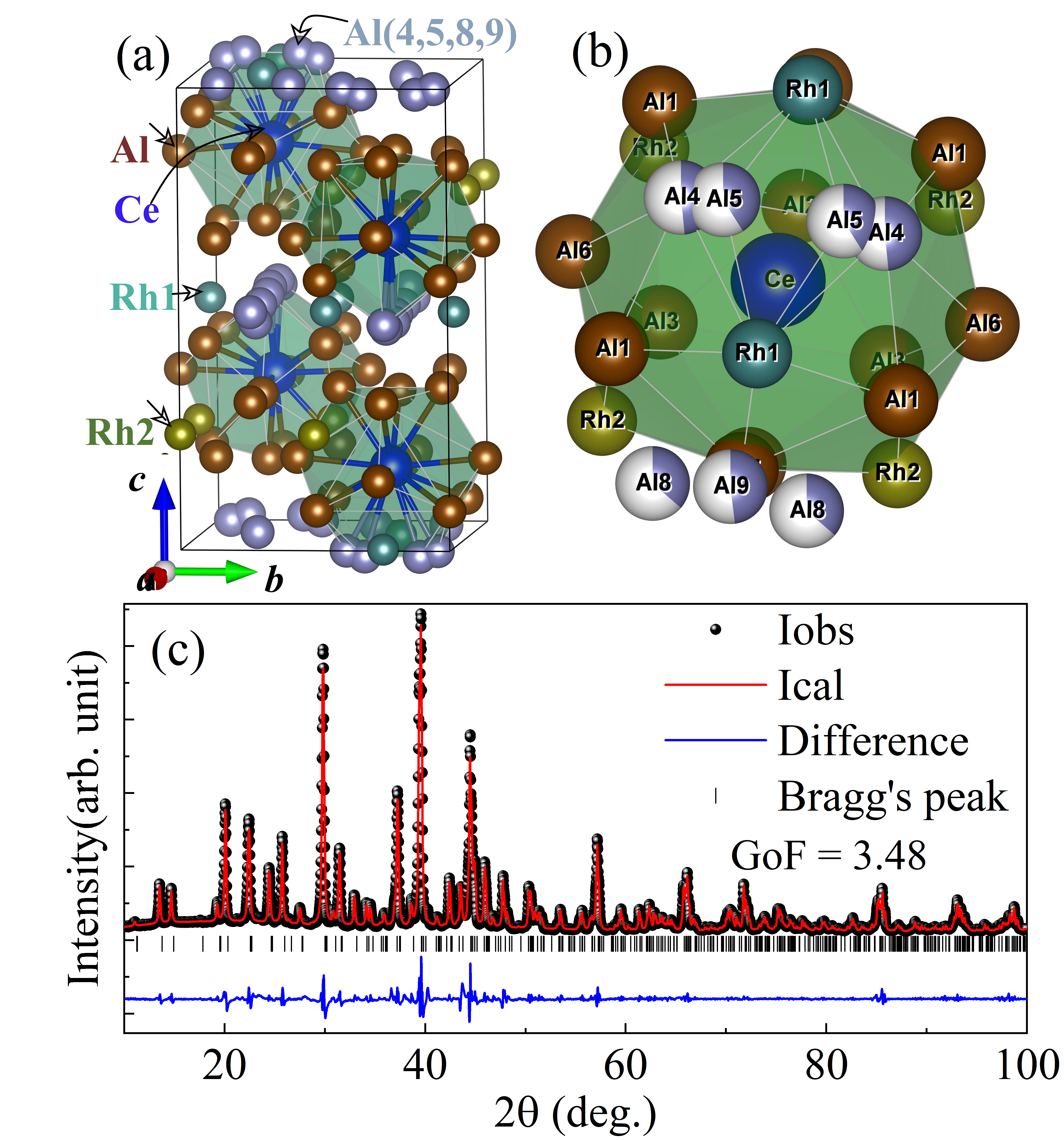}
		\caption{(a) The crystal structure of CeRh$_4$Al$_{15}$. The blue, dark yellow, and wine color atoms are Ce, Rh, and Al, respectively. (b) The caged type local environments of Ce. showing two Rh sites and nine Al sites. The Al sites (Al4, Al5, Al8, Al9) are partially occupied. (c) Room temperature Rietveld fitted XRD pattern of CeRh$_4$Al$_{15}$.}
		\label{XRD}
\end{figure}

\begin{table*}
	\centering
	\caption {Crystallographic parameters of CeRh$_4$Al$_{15}$ with NdRh$_4$Al$_{15}$-type tetragonal structure (space group $P4_2/nmc$) obtained from a single-crystal x-ray analysis and a Rietveld refinement of powder XRD.}
	\vskip .5cm
	\addtolength{\tabcolsep}{+1.5pt}
	\begin{tabular}{c c c c c c c c}
		\hline
		\hline
		
		Atomic coordinates &  	& &&&     \\[1ex]
		
		Atoms &  $x$         & $y$         &   $z$          & occ.      & $U$         & site   & Sym. \\[1ex]
		Ce   & 1/4 	    & 1/4         &   0.33506(2)   & 1         & 0.00571(6)  & $4d$   & 2$mm$  \\[1ex]
		Rh1  & 1/4	    & 0.01207(4)  &   0.49810(2)   & 1         & 0.01091(6)  & $8g$   & .$m$.  \\[1ex]
		Rh2  & 0.50167(2)   & 0.49833(2)  &   1/4          & 1         & 0.00544(6)  & $8f$   & ..2  \\[1ex]
		Al1  & 0.00326(10)  & 0.02576(12) &   0.41532(5)   & 1         & 0.01098(13) & $16h$  &  1   \\[1ex]
		Al2  & 1/4	    & 0.01350(13) &   0.18307(8)   & 1         & 0.00870(16) & $8g$   & .$m$.  \\[1ex]
		Al3  & 1/4	    & 0.05928(16) &   0.66241(7)   & 1         & 0.01093(18) & $8g$   & .$m$.  \\[1ex]
		Al4  & 1/4	    & 0.0805(5)   &   0.01655(18)  & 0.580(5)  & 0.0204(5)   & $8g$   & .$m$.  \\[1ex]
		Al5  & 1/4	    & 0.1478(6)   &   0.0289(3)    & 0.420(5)  & 0.0204(5)   & $8g$   & .$m$.  \\[1ex]
		Al6  & 1/4	    & 0.59918(12) &   0.81372(7)   & 1         & 0.00817(15) & $8g$   & .$m$.  \\[1ex]
		Al7  & 1/4	    & 0.59993(13) &   0.34405(8)   & 1         & 0.01069(17) & $8g$   & .$m$.  \\[1ex]
		Al8  & 1/4	    & 0.6185(4)   &   0.0054(3)    & 0.338(8)  & 0.0158(10)  & $8g$   & .$m$.  \\[1ex]
		Al9  & 3/4	    & 1/4         &   0.0232(2)    & 0.712(13) & 0.0222(9)   & $4c$   & 2$mm$  \\[1ex]
			\hline
		Lattice parameters 	&  &&&&	   \\[1ex]
		                & single-crystal study & & & & & Rietveld refinemnet	  \\[1ex]
		$a$(\AA) 	& $c$(\AA) & $V$(\AA$^3$) &&&	$a$(\AA) 	& $c$(\AA) & $V$(\AA$^3$)   \\[1ex]
		
		9.1282(4) &  15.6305(6) & 1302.40(12) &&&    9.134(2) &  15.605(3) & 1301.9(5)	  \\[1ex]
			\hline
		Refinement quality & &&&& 	    \\[1ex]
		        & single-crystal study & & & & & Rietveld refinemnet	  \\[1ex]

		$R[F^2 > 2\sigma (F^2)]$ & $wR(F^2)$ & $S$  &&&	 $\chi^2$ & $R_P$(\%) &$R_{WP}$(\%)   \\[1ex]
		
	    0.0436	 &  0.1107  &  1.234 &&& 1.67	 &  0.5   &   0.4   \\[1ex]
		
		\hline
		\hline
	\end{tabular}
	\label{table1}
\end{table*}

\section{EXPERIMENTAL METHODS}
\label{expt}

The polycrystalline samples of CeRh$_4$Al$_{15}$ and LaRh$_4$Al$_{15}$ were prepared according to Ref.~\cite{NESTERENKO20191061}. Powder x-ray diffraction (PXRD) with Mo-$K_\alpha$ radiation ($\lambda$ = 0.71073 \AA; 50~kV; 20~mA) was used to determine the phase purity and crystal structure. Here, we prepared high-quality single crystal of CeRh$_4$Al$_{15}$ using Al-flux method. A starting composition of Ce$_{1}$Rh$_4$Al$_{30}$ was used and the elements were placed in an alumina crucible then sealed in a quartz tube with 1/3 atm of pure Ar gas. The quartz ampoule was cooled slowly from 1150~$^{\circ}$C to 900~$^{\circ}$C at a rate of 1~$^{\circ}$C/hr, then cooled quickly to 750~$^{\circ}$C before centrifugation in order to avoid formation of secondary phases.
For the crystal structure of CeRh$_4$Al$_{15}$, a single crystal x-ray analysis is also performed by using a small piece (89 $\mu$m $\times$ 72 $\mu$m $\times$ 69 $\mu$m) prepared by crushing a single crystal. Diffraction data were collected at room temperature on a RIGAKU AFC11 Saturn 724+ CCD diffractometer and a Mo K$_{\alpha}$ radiation monochromated by a VariMax confocal X-ray optics device. Structure solution and refinement were carried out using the SHELXL program (version 2018/3)~\cite{Sheldrick2015}. The obtained atomic coordinates were standaridized using the STRUCTURE TIDY program~\cite{STRUCTURETIDY}.

DC magnetization of a single crystalline sample was measured down to 2~K and in high fields up to 7~T using a Quantum Design Magnetic Property Measurement System (MPMS-SQUID) with the applied magnetic field along the crystallographic $c$ and $a$-axes. The specific heat measurements were carried out using a Physical Property Measurement System (PPMS) with an adiabatic demagnetization refrigerator (ADR) option down to 70~mK (adiabatic heat-pulse method) in zero-field and using a He-3 system down to 0.4 K in various applied magnetic fields (relaxation method). Electrical resistivity measurements in the temperature range 2~K to 300~K (AC 4-terminal method) and 80~mK to 2~K (DC 4-terminal method) were performed using a GM refrigerator and Quantum Design PPMS with ADR option, respectively  with the current along the crystallographic $c$ and $a$-axes and magnetic field in the $ab$ plane.

Zero-field (ZF) and longitudinal-field (LF)-$\mu$SR measurements on the polycrystalline sample of CeRh$_4$Al$_{15}$ were performed at the ISIS Neutron and Muon Source, UK, using the EMU spectrometer. For ISIS muon measurements, a crushed polycrystalline sample was mounted on a 99.999\% pure silver plate using diluted GE varnish to ensure good thermal contact and then covered with a thin silver foil. We used a dilution fridge to cool the samples down to 70~mK. The $\mu$SR data were analyzed with the MANTID~\cite{ARNOLD2014156} and Wimda software~\cite{PRATT2000710}. 

The INS experiments on polycrystalline of CeRh$_4$Al$_{15}$ and LaRh$_4$Al$_{15}$ (phonon reference) were performed on the MERLIN time of flight (TOF) spectrometer at the ISIS Neutron and Muon source, UK~\cite{Bewley2009}. The powdered samples of these materials were filled in a thin Al-foil envelope and mounted in an annular form inside thin-walled cylindrical Al-cans with a diameter of 30~mm and height of 40~mm. Low temperatures down to 5~K were obtained by cooling the sample mounts in a top-loading closed cycle refrigerator with He-exchange gas. The INS data were collected with repetition-rate multiplication using a neutron incident energy of $E_i$ = 75~meV, and a Fermi chopper frequency of 250~Hz, which also provided data for $E_i$ = 24.7 and 12~meV. The elastic resolution (FWHM) was 5.1~meV for $E_i$ = 75~meV, 0.97~meV for $E_i$ = 24.6~meV and 0.37~meV for $E_i$ = 12~meV. The data are presented in absolute units, mb/(meV sr f.u.) using the absolute normalization obtained from the standard vanadium sample measured in identical conditions.

\section {Experimental results and discussion}


\begin{figure*}
		\includegraphics[width=\textwidth, keepaspectratio]{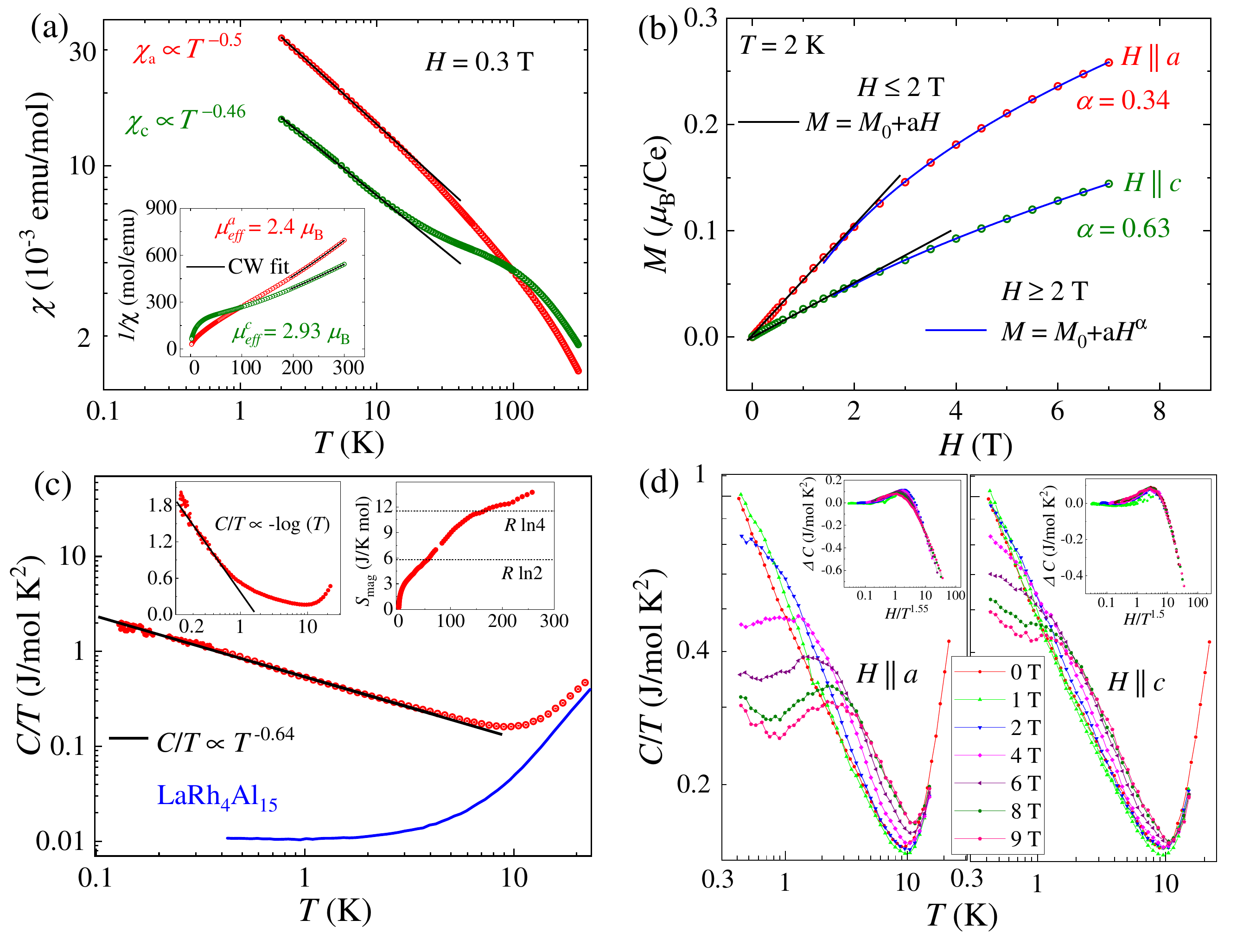}
		\caption{(a) Log-log plot of the magnetic susceptibility vs. temperature of CeRh$_4$Al$_{15}$ under applied magnetic field along the $c$ and $a$-axes. The solid lines are fits to the data with $\chi \propto T^{-1+\alpha_{\chi}}$ behavior. The inset shows inverse susceptibility with the solid lines representing Curie-Weiss behavior at high temperatures. (b) Magnetization as a function of field along $c$ and $a$-axes at $T = 2$~K, showing a linear behavior, $M \sim H$, for $H <$ 2~T and a power-law behavior, $M = M_0 + H^{\alpha}$, for $H >$ 2~T. (c) Low-temperature specific heat of CeRh$_4$Al$_{15}$ and LaRh$_4$Al$_{15}$ on log-log plot. The solid line is fit to the data with $C/T \propto T^{-1+\alpha_{\text{C}}}$ behavior. The left inset shows $C/T$ vs log $T$ and is fitted by $C /T \sim$ - log $T$ (solid line). The right inset shows the magnetic ($4f$) entropy ($S_{\text{mag}}$) as a function of temperature obtained from the heat capacity data as described in the text. (d) Low-temperature specific heat of CeRh$_4$Al$_{15}$ (on a log-log scale) at varying fields applied parallel to the $c$ and $a$-axes. The insets show $\Delta C $ = $C$($H$)/$T$ - $C$(0)/$T$ vs. $f$($H$/$T^{\beta}$) scaling in the temperature range 0.4 $\leq T \leq$10~K.}
		\label{transport}
\end{figure*}

The crystal structure and PXRD pattern of CeRh$_4$Al$_{15}$ are presented in Fig.~\ref{XRD}(a-b) and Fig.~\ref{XRD}(c), respectively. The Rietveld refinement of the XRD pattern of CeRh$_4$Al$_{15}$ presented in Fig.~\ref{XRD}(c) reveals the single phase nature of the polycrystalline sample. Powder and single crystal x-ray diffraction studies confirm that CeRh$_4$Al$_{15}$ and LaRh$_4$Al$_{15}$ compounds crystallize in the NdRh$_4$Al$_{15}$-type tetragonal structure with space group $P4_2/nmc$. The obtained lattice parameters, atomic position parameters, and thermal parameters shown in Table~\ref{table1} are in good agreement with the values reported in the previous literature~\cite{NESTERENKO20191061}.
As shown in Fig.~\ref{XRD}(a-b), the Ce atoms are surrounded by six Rh and 14 Al atoms which results in a distorted prismatic hexagonal environment of Ce atom with eight additional atoms capping all faces of the prism Ce[Rh$_4$Al$_{16}$]. These “cages” are each bounded by sharing four common vertices (Rh2) into a corrugated slab running perpendicular to the $c$-axis.

\begin{figure}
\includegraphics[width=8.0 cm, keepaspectratio]{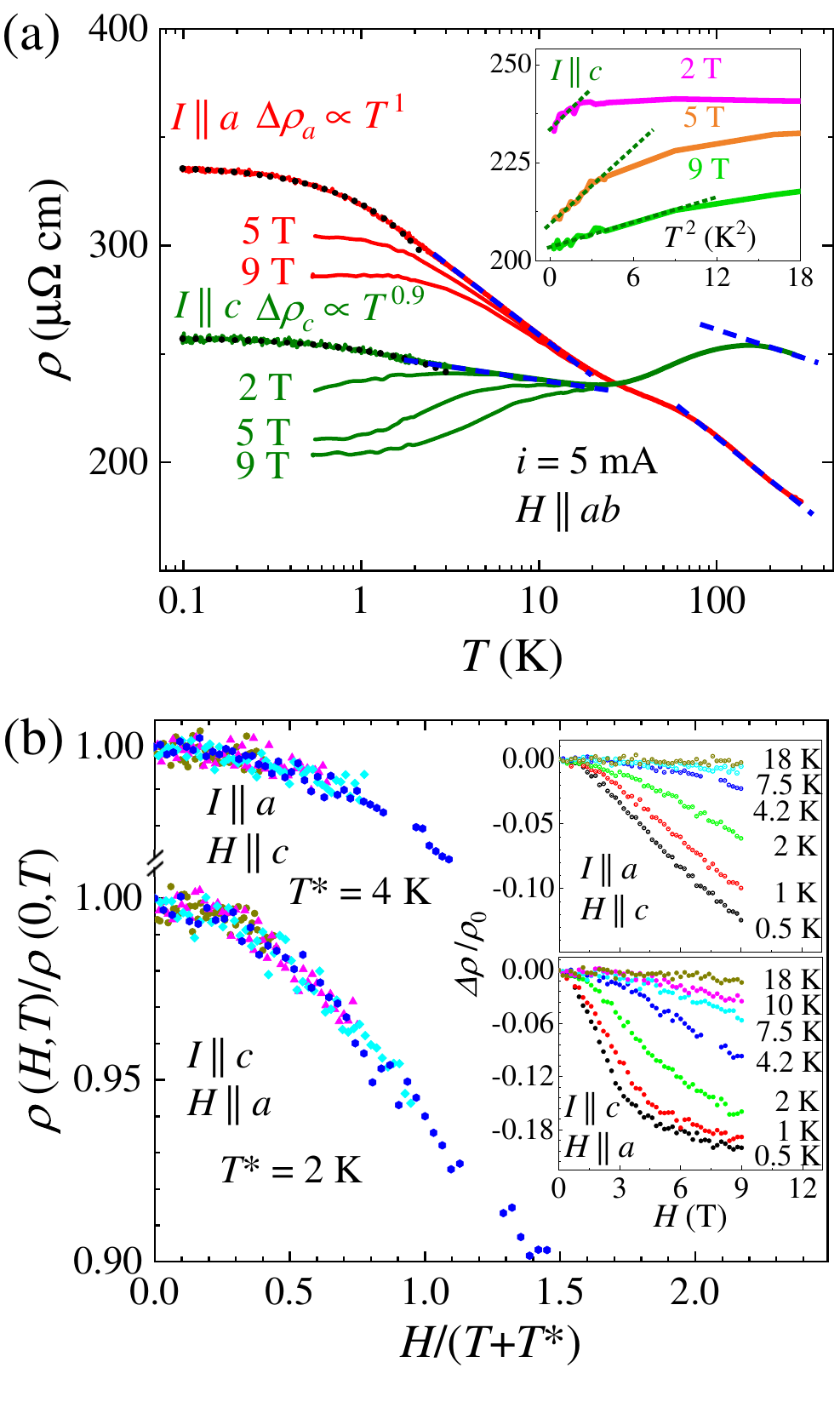}
\centering
\caption{ (a) Temperature dependence of the electrical resistivity and its magnetic field response down to 0.1~K with the current flowing along $a$ and $c$-axes and the magnetic field applied within the crystallographic $ab$ plane on a log scale. It is to be noted that there is a factor 10 missing in the absolute value of resistivity of the polycrystalline sample of CeRh$_4$Al$_{15}$ reported in ref.~\cite{NESTERENKO20191061}. The dotted black lines represent the power law fit as discussed in the text. The blue dash lines represent the -log($T$) behavior. The inset shows resistivity vs $T^2$ plot in various applied magnetic fields for the current flowing along the $c$-axis, indicating the validity temperature range of the $T^2$ law. (b) Normalized resistivity is plotted as a function of $H$/($T + T^*$), where $T^*$ is the characteristic temperature, show the scaling behavior in the temperature range 4~K to 18~K. Insets: The normalized resistivity plotted as a function of applied magnetic field for $I\parallel a$ and $H\parallel c$ (upper Inset), and $I\parallel c$ and $H\parallel a$ (lower Inset).}
\label{MR}
\end{figure}

According to an earlier structural investigation~\cite{NESTERENKO20191061}, CeRh$_4$Al$_{15}$ contains twelve crystallographic sites, including one cerium site, two rhodium sites, and nine aluminium sites with four of them are partially occupied; Al4 splits to Al4 and Al5, and Al9 splits to two Al8 and Al9. Moreover, these unoccupied Al sites are arrayed parallel to the tetragonal $a$-axis in Rh2-Al layer. As per the reference, these four sites are not shared with any other atoms which means that these are not due to disorder but occupied partially.

We tried to refine the data by manually adding the disorder at a few sites. However, the refinement either diverges or doesn't change significantly with disordered structure. Based on the Rietveld refinement, it is hard to exclude the possibility of disorder or site mixing between Al and Rh sites. We have also checked the chemical composition of single-crystal CeRh$_4$Al$_{15}$ by an electron probe microanalysis (EPMA, not shown here), which shows very good agreement between those obtained by XRD (CeRh$_4$Al$_{15.39}$) and EPMA (CeRh$_{3.98}$Al$_{15.43}$).
Apparently, the disorder induced by Al atoms with partial occupancy is an intrinsic feature with NdRh$_4$Al$_{15.40}$-type structure. In addition, all interatomic distances are of two different groups. The first one includes Ce-Ce, Ce-Rh, and Rh-Rh are close to or slightly smaller than the sum of the respective metallic single-bond radii and thereby could be regarded as chemical bonding.
The second group includes Ce-Ce ($>$6.965 \AA), Ce-Rh ($>$3.337 \AA), and Rh-Rh ($>$4.316 \AA) distances significantly exceeding the sum of metallic radii of 3.66 \AA, 3.18 \AA, and 2.69 \AA, respectively, clearly indicating a different environment of magnetic Ce site.

The magnetic susceptibility measured under the applied magnetic field along $c$ and $a$-axes are shown in Fig.~\ref{transport}(a). As is apparent from the inset of Fig.~\ref{transport}(a) that the data above 100~K along both directions are best described by modified Curie-Weiss law with the parameters $\chi^a_0$ = 5~$\times$ 10$^{-4}$~emu/mol, $\mu^a_{\text{eff}}$ = 2.4~$\mu_{\text{B}}$, $\theta^a_{\text{P}}$ = -74~K for $H\parallel a$ and $\chi^c_0$ = 6~$\times$ 10$^{-4}$~emu/mol, $\mu^c_{\text{eff}}$ = 2.9~$\mu_{\text{B}}$, $\theta^c_{\text{P}}$ = -139~K for $H\parallel c$. The effective magnetic moments are comparable and slightly larger than the free Ce$^{3+}$ ion value 2.54~$\mu_{\text{B}}$ for the applied magnetic field along $a$ and $c$-axes, respectively. The sign of the Weiss temperature suggests AFM correlations of the Ce moments. The susceptibility is very anisotropic along the two different crystallographic directions. At the lowest temperature, it reaches a two-times high value for an applied magnetic field along the basal plane as compared to the susceptibility with an applied field along the $c$-axis. This observation suggests that the anisotropic AFM spin fluctuations are much stronger along the basal planes. The interesting observation is the cross-over seen in the temperature dependence of the susceptibility of $a$- and $c$-axes near 100~K indicating the change from an easy $c$-axis to an easy plane anisotropy.

Upon cooling, $\chi$($T$) exhibits a power-law behavior over nearly a decade in temperature $2 \leq T \leq$ 10~K, as $\chi$($T$)= $a\times T^{-1+\alpha_{\chi}}$ with $\alpha_{\chi}$ = 0.5 and 0.54 for $H$ applied in the basal plane and along the $c$-axis, respectively, with no evidence of magnetic order down to 2~K. This type of power-law behavior of $\chi$($T$) is observed for several NFL systems such as CeRhBi~\cite{SASAKAWA2005111,doi:10.7566/JPSJ.87.064708,PhysRevB.101.214437} and suggests CeRh$_4$Al$_{15}$ is located near a QCP. 

On the other hand, the magnetization curves at 2~K are linear up to 2~T followed by a curvature above 2~T up to the highest available field of 7~T and are best described by power-law behavior $M$($H$) = $M_0$ + $aH^{\alpha}$, as shown in Fig.~\ref{transport}(b). Here $M_0$ is the negative constant offset and $\alpha$ is the same as in the power exponent of $\chi$, i.e., $\alpha_{\chi}$. The power-law fit to the high field magnetization data (2–7~T) yields the exponent $\alpha_{\text{a}}$ = 0.34(1) and $\alpha_{\text{c}}$ = 0.63(4) for the applied magnetic field along $a$ and $c$-axes, respectively. Many systems close to a QCP exhibit similar behavior and are attributed to a Griffiths phase scenario~\cite{PhysRevLett.118.267202,PhysRevB.99.224424}. We also fit the magnetization isotherm data to the modified Langevin function (not shown here), which also explains the behavior satisfactorily, indicating the paramagnetic ground state at 2 K.

We further performed zero-field heat capacity measurements on the single crystal sample of CeRh$_4$Al$_{15}$ as shown in Fig.~\ref{transport}(c). The low temperature heat capacity diverges with decreasing temperature with a power-law-like behavior $C$($T$)/$T$ = $a\times T^{-1+\alpha_{\text{C}}}$ with $\alpha_{\text{C}}$ = 0.44 in the temperature range 0.1 $\leq T \leq$ 7~K. Similar power-law behavior in the heat capacity has been reported for some chemically substituted systems exhibiting NFL behavior with strong disorder near a critical concentration, also called Griffiths singularities, e.g., Ce$_{1-x}$Th$_x$RhSb~\cite{PhysRevB.49.348}, Ce(Cu$_{1-x}$Co$_x$)$_2$Ge$_2$~\cite{PhysRevB.98.165136}. Similar power-law behavior in the low temperature heat capacity has also be seen for some stoichiometric NFL systems like CeInPt$_4$~\cite{PhysRevB.76.174439}. Here it is to be noted that our data diverges more strongly than the standard $C/T\sim$ -log$T$ behavior found in many heavy fermion systems displaying NFL behavior near an AFM QCP~\cite{PhysRevLett.72.3262,RevModPhys.73.797}. It can be seen in the left inset of Fig.~\ref{transport}(c) that $C$/$T$ is roughly proportional to - log $T$ over a narrow temperature range from 0.13 to 0.7~K. Our data are also in marked contrast to the $C_{\text{4f}}$ /$T$ $\propto$ 1 - $a\sqrt{T}$, predicted by the spin-fluctuation theory at the AFM QCP in 3D~\cite{RevModPhys.79.1015, RevModPhys.73.797}. Furthermore, the calculated heat capacity of the NFL system  UCu$_{4}$Ni using the Kondo disorder model exhibits $C$/$T$ $\propto$ - log $T$ ~\cite{doi:10.1063/1.373270}, which is not the case for CeRh$_4$Al$_{15}$, and hence we rule out the possibility of Kondo disorder model for  CeRh$_4$Al$_{15}$.

The electronic coefficient of specific heat $\gamma$ = 23(6)~mJ/mol K$^2$ is obtained from a linear fitting of the $C$/$T$ versus $T^2$ data in the range of 10 $\leq T \leq $ 15~K (fit not shown), reflecting the moderate heavy fermion behavior in CeRh$_4$Al$_{15}$. With $\gamma$ = 10.0(5)~mJ/mol K$^2$ for LaRh$_4$Al$_{15}$, we estimated the renormalized quasi-particle mass in CeRh$_4$Al$_{15}$ [$\gamma$(CeRh$_4$Al$_{15}$)/$\gamma$(LaRh$_4$Al$_{15}$) $\sim$ 2.3] to be $m^*$ $\sim$ 2.3 $m_e$, where $m_\text{e}$ is the free electron mass.

The magnetic entropy $S_{\text{mag}}$, calculated by integrating the $C_{\text{4f}}$/$T$ versus $T$, where $C_{\text{4f}}$ = $C$(CeRh$_4$Al$_{15}$) - $C$(LaRh$_4$Al$_{15})$, is shown in the right inset of Fig.~\ref{transport}(c). The value of entropy is $0.3R$ln2 at $T = 4$~K and $R$ln2 at 56~K. The reduced value of magnetic entropy suggests the presence of Kondo screening of the $f$ moment by the conduction electrons. The full entropy expected for the $J = 5/2$ multiplet of Ce$^{3+}$ is recovered at room temperature, which indicates that overall the crystal electric ﬁeld (CEF) splitting energy is close to 300~K.

Figure~\ref{transport}(d) shows the variation of log $C/T$ vs log $T$ with the fields $H \leq 9$~T applied parallel to the $c$ and $a$-axes. The low temperature upturn for $H = 0$ transforms into a maximum with increasing $H$. The data for $H > 0$ show a broad maximum, which shifts to higher temperatures with increasing $H$ typical behavior of Kondo systems under magnetic ﬁeld. The evolution of the Schottky specific heat results from the splitting of the CEF ground state doublet by the Zeeman effect via the excited Ce$^{3+}$ ($4f^1, J = 5/2$) state that splits into three Kramers doublets. The change in the low temperature heat capacity is larger for the magnetic field parallel to the basal plane, than the field along the $c$-axis, leading to a gradual recovery of the properties of a FL, in agreement with $\rho_a$ ($T$) discussed in the next section. A very similar behavior has been reported in other NFL compounds~\cite{PhysRevLett.82.1293}.

The scaling of specific heat as a function of magnetic field can provide further insight into the origin of the observed NFL behavior~\cite{PhysRevLett.67.2886,PhysRevB.48.9887}.
The insets of Fig.~\ref{transport}(d) show that, in the temperature range 0.4 $\leq T \leq$10~K, our data are consistent with the, scaling relation $\Delta C $ = $C$($H$)/$T$ - $C$(0)/$T$ = $f$($H$/$T^{\beta}$) with $\beta$ = 1.5 and 1.55 for the field applied parallel to the $c$ and $a$-axes, respectively. Because an exponent $\beta$ greater than 1 indicates a non-single impurity effect~\cite{PhysRevLett.67.2886}, our scaling analysis rules out a single-ion effect as a possible reason for the NFL behavior and may be taken as further evidence of the Griffiths phase as a possible origin for the NFL behavior.

Figure~\ref{MR}(a) shows the temperature evolution of the electrical resistivity and its magnetic field response from 0.1 -- 300~K with the current flowing along $a$ ($\rho_a$) and $c$ ($\rho_c$)-axes and the magnetic field applied within the crystallographic $ab$ plane. The resulting $\rho_a$ and $\rho_c$ are highly anisotropic with similar behavior along both directions. For both directions at high temperatures, $\rho$ exhibits - log($T$) behavior [blue dash lines in Fig.~\ref{MR}(a)] followed by a broad maximum. Below these maxima, both $\rho_a$ and $\rho_c$ keep increasing logarithmically with decreasing temperature followed by a saturation-like behavior at low temperatures. The overall behavior of the resistivity can be explained based on the Kondo effect in the presence of the CEF as proposed by the theory of Cornut and Coqblin, who treated the problem of resistivity of single Kondo impurities in the presence of CEF~\cite{PhysRevB.5.4541}. It is important to note that at room temperature $\rho_a$ is significantly smaller than $\rho_c$ and the anisotropy switches at a lower temperature, which is consistent with the magnetic susceptibility and can be explained based on the coupling of the conduction electrons with the components of the magnetic moment (i.e. larger component of the moment will give stronger scattering).

In zero-field, a $T^2$ dependence of the low-temperature resistivity, characteristic of FL behavior, is absent in the two principal directions, instead, $\rho$($T$) varies linearly with temperature ($\Delta \rho$($T$) $\sim$ $T^{\epsilon}$; $\epsilon_a =1$ and $\epsilon_c=0.9$)  [Fig.~\ref{MR}(a)] and is characteristic of an NFL state~\cite{PhysRevLett.81.1501,PhysRevLett.85.626,SRJulian_1996,PColeman_2001}. Moreover, the effect of the magnetic field has a very interesting effect on the low temperature resistivity. With increasing magnetic field, the low temperature diverging resistivity is lost only for $\rho_c$ ($H \parallel ab$ with the current along $c$ direction) [Fig.~\ref{MR}(a)], and the temperature dependence follows a $T^2$ law, like a typical FL [inset of Fig.~\ref{MR}(a)]. The range of temperatures in which FL behavior is observed, increases with increasing field, which is similar to that observed in the stoichiometric compound CeNi$_2$Ge$_2$~\cite{schlottmann1983bethe}. However, $\rho_a$ saturates similar to the zero-field case even in a field of 9~T, though with a decreased saturation value.

The insets of Fig.~\ref{MR}(b) show the magnetoresistance (MR) $\rho$($H$) vs magnetic field at different $T$ for current parallel to the $a$- ($H\parallel c$) and $c$-directions ($H\parallel a$). The MR for both directions are negative over the entire temperature range of $0.55 \leq T \leq 18$~K. In the paramagnetic region, the negative MR is due to the freezing out of spin-flip scattering in a Kondo compound by the magnetic field. Whereas, the positive MR in the ordered state is consistent with the AFM nature of the magnetic ordering~\cite{PhysRevB.90.155101}. Therefore, the behavior of the negative MR could be attributed to the absence of magnetic ordering in the Kondo lattice system CeRh$_4$Al$_{15}$. In order to estimate $T_{\mathrm{K}}$, the normalized MR $\rho(H) / \rho(0)$ plotted as a function of $H$/($T + T^*$), is presented in Fig.~\ref{MR}(b). This allows us to scale the MR data measured at different temperatures (from 4~K to 18~K ) onto a single curve. Here, $T^*$ is the characteristic temperature which is an approximate measure of the $T_{\mathrm{K}}$~\cite{PhysRevB.62.8950}. Thus estimated values of $T_{\mathrm{K}}$ = 4 and 2~K for $I\parallel a$; $H\parallel c$ and $I\parallel c$; $H\parallel a$ respectively, are in excellent agreement with $T_{\mathrm{K}}$ = 3.1(3)~K reported for the polycrystalline sample~\cite{NESTERENKO20191061}. 

All these results confirm the NFL ground state of CeRh$_4$Al$_{15}$ single crystal and lead to further examination using a microscopic technique such as muon spin relaxation measurements.

\begin{figure*}
\begin{center}
		\includegraphics[width=\textwidth, keepaspectratio]{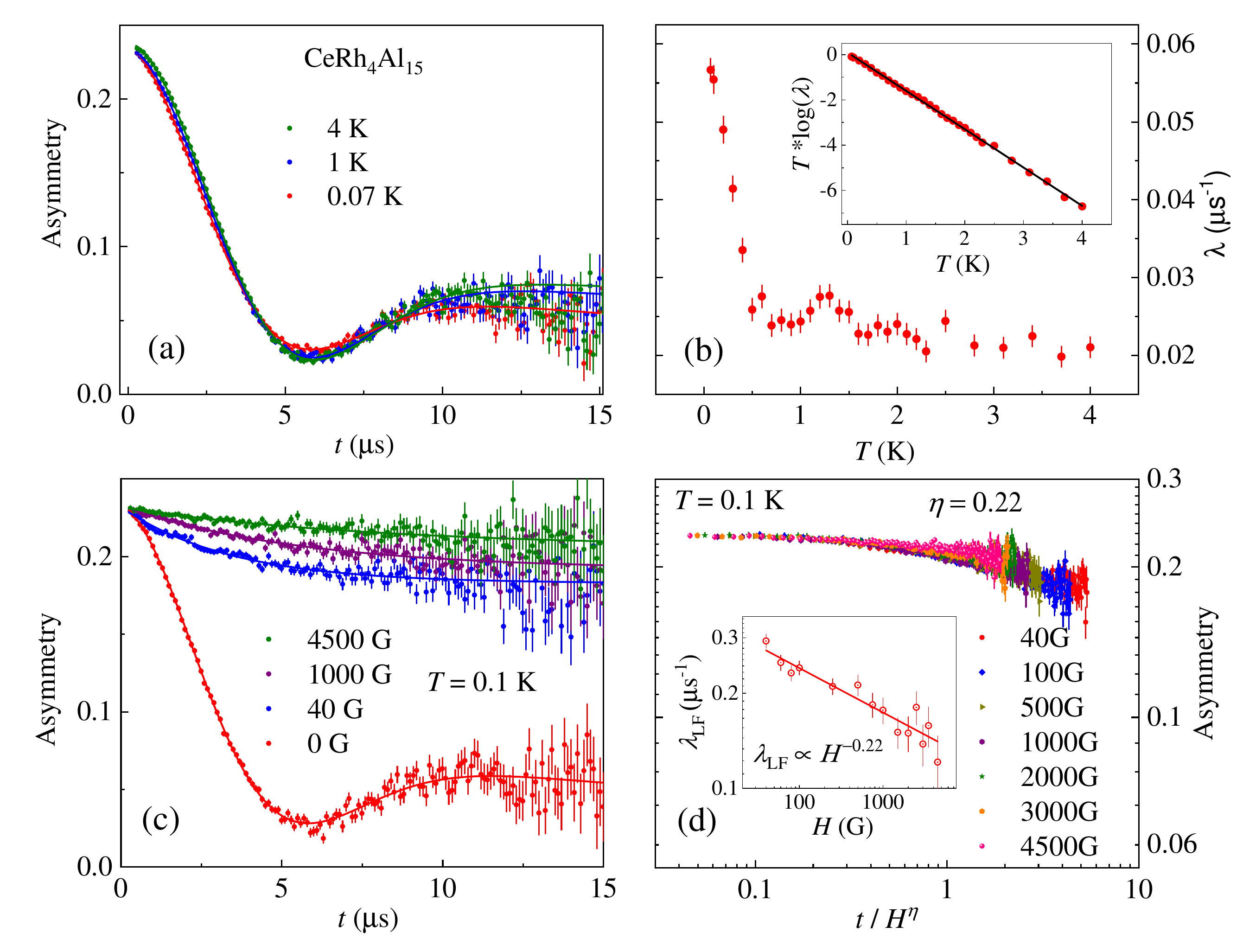}
		\caption{(a) ZF-$\mu$SR spectra of CeRh$_4$Al$_{15}$ at representative temperatures. The solid lines are the fitted curves (see the text for details). (b) Temperature dependence of the muon spin relaxation rate ($\lambda$). The inset described an activation like behavior of $\lambda$, i.e., $\lambda$($T$) = $\lambda_0 \exp$(- $E_\mathrm{g}$/$k_\mathrm{B} T$), over a given temperature range. (c) LF-$\mu$SR spectra at 0.1~K at different fields up to 4500~G. The solid lines are the fitted curves (see the text for details) (d) Time-field scaling of the asymmetry function $A$($t$) vs $t/H^{\eta}$ at 0.1~K. The inset shows the log-log plot of $\lambda_{LF}$ vs $H$ with a solid line representing the power-law behavior.}
		\label{ZF_spectra}
		\end{center}
	\end{figure*}

To further probe the nature of the ground state observed from the thermal and transport measurements at low $T$, we carried out ZF- and LF-$\mu$SR measurements down to 70~mK. $\mu$SR is a powerful local probe that is able to detect tiny magnetic moments with an average ordered moment size of 0.005~$\mu_B$ (or higher) and can distinguish the random static fields associated with, for example, the dipolar coupling of the muon and quasi-static nuclear moments and dynamically fluctuating fields associated with electronic spin fluctuations. Details of the $\mu$SR technique can be found in Ref.~\cite{MUSR}. The time-dependent ZF-$\mu$SR spectra of CeRh$_4$Al$_{15}$ collected at various temperatures between 0.07 -- 4~K are displayed in Fig.~\ref{ZF_spectra}(a). At high temperatures, the $\mu$SR spectra show faster depolarization, a minimum near 5 $\mu$s and recovery at a higher time, which is a typical behavior arising from the nuclear moment contribution. When cooling below 1~K a weak temperature dependence of relaxation rate is observed, which is due to electronic relaxation. However, neither the oscillatory signal nor a 2/3 loss of the initial asymmetry of the muon polarization is observed. This behavior suggests the absence of a well-defined or disordered static magnetic field (from electronic moments) at the muon stopping site and hence ruled out any possibilities of long-range magnetic ordering or spin freezing due to Ce$^{3+}$ moments.  

The ZF-$\mu$SR spectra have been successfully fitted to the muon spin relaxation function described below.
\begin{equation}
A_z(t) =  A_0 G_{\mathrm{KT}}(t)\times \exp(-\lambda t) + A_{\mathrm{BG}}
\end{equation}

\noindent where $A_0$ is the initial asymmetry, $\lambda$ is the muon spin relaxation rate accounting for the dynamic magnetic fields due to fluctuating electronic moments, $A_{\mathrm{BG}}$ is a constant background arising from muons stopping on the silver sample holder, and $G_{\mathrm{KT}}$ is the static Kubo–Toyabe function describing the muon spin depolarization with a rate of $\sigma_{\mathrm{KT}}$ caused by randomly oriented $^{103}$Rh and $^{27}$Al nuclei and is given by ~\cite{PhysRevB.20.850}

\begin{equation}
G_{\mathrm{KT}}(t) = \frac{1}{3}+\frac{2}{3}[1-(\sigma_{\mathrm{KT}}t)^2]\exp(\frac{-\sigma_{\mathrm{KT}}^2t^2}{2})
\end{equation}

The fits to the spectra by Eq. (1) are shown by the solid curves in Fig.~\ref{ZF_spectra}(a). The fitting parameter $\lambda$ determined from the best fits is displayed in Fig.~\ref{ZF_spectra}(b). The value of $\sigma_{\mathrm{KT}}$ was obtained to be $\sim$ 0.295 $\mu s^{-1}$ from fitting the spectra at 4~K, and this value was found to be nearly temperature independent down to 70~mK. The $A_{\mathrm{BG}}$ (= 0.02) was estimated from 70~mK data and was kept fixed for fitting all the other spectra. 

It can be seen from Fig.~\ref{ZF_spectra}(b) that $\lambda$ exhibits a sharp increase as $T$ decreases, without any sign of static magnetic ordering down to 70~mK. The $\lambda$ shows an activated behavior $\lambda$($T$) = $\lambda_0 \exp$(-$E_\mathrm{g}/k_\mathrm{B}T$) or equivalently $T$ log($\lambda$) = $T$ log($\lambda_0$)--$E_\mathrm{g}/k_\mathrm{B}$, where $E_\mathrm{g}$ is the energy gap and $k_\mathrm{B}$ is the Boltzmann’s constant. A linear fit of $T$ log($\lambda$) versus $T$ plot [inset of Fig.~\ref{ZF_spectra}(b)] yields $E_\mathrm{g}$ = 90(1)~mK. Thus the spin dynamics seem to be thermally activated, indicating the presence of low energy spin fluctuations in CeRh$_4$Al$_{15}$. Very similar behavior was also observed for the stoichiometric NFL systems CeInPt$_4$ and CeRhBi with an activation energy of 2.9~mK and 140(3)~mK, respectively~\cite{PhysRevB.76.174439,doi:10.7566/JPSJ.87.064708}.

We further carried out LF-dependent measurements in order to determine the dynamics of the electronic magnetic moment fluctuations in CeRh$_4$Al$_{15}$. When a small LF (about 25~G) is applied, the observed weak contribution from the nuclear magnetic moments observed in the ZF signal is eliminated. On the other hand, a large LF is needed to decouple the muon depolarization from the internal field arising from the fluctuating electronic spins. The representative LF-spectra are displayed in Fig.~\ref{ZF_spectra}(c). It is seen that even with 4500~G LF it is not sufficient to suppress the muon relaxation at 0.1~K completely. This reveals that the magnetic ground state is entirely dynamic at the base temperature. However, field dependent spectra at $T =$4~K (not shown here) behave as expected for the high-temperature paramagnetic state. The LF spectra measured at 0.1~K under several magnetic fields ($\geq$ 40 G) can also be modeled by Eq. (1) with $G_{\mathrm{KT}}$($t$) = 1. The obtained $\lambda_{\mathrm{LF}}$ as a function of the field is shown in the inset of Fig.~\ref{ZF_spectra}(d). The variation of the $\lambda_{\mathrm{LF}}$($H$) can be represented by a power-law behavior with a power exponent -0.22(3).

Furthermore, our LF-$\mu$SR data follow characteristic time-field scaling $A(t,H)$ = $A(t/H^\eta)$, where the exponent $\eta$ provides information about spin–spin dynamical autocorrelation~\cite{PhysRevLett.75.2023,PhysRevLett.87.066402,PhysRevB.20.850,MacLaughlin_2004}. By observing the time-field scaling, independent information on the nature of the spin autocorrelation function $q$($t$) = $\langle S_i$($t$)$S_i$(0)$\rangle$ can be obtained. In an inhomogeneous system, $q$($t$) is theoretically predicted to exhibit a power law behavior for $\eta < 1$ and stretched exponential behavior for $\eta >$ 1~\cite{MacLaughlin_2004,PhysRevLett.77.1386}. The asymmetry as a function of scaling variable $t/H^\eta$ at 0.1~K is shown in Fig.~\ref{ZF_spectra}(d). The best scaling of the overall data can be obtained with $\eta$ = 0.22, for the $H_{\mathrm{LF}}$ data up to 4500~G. The scaling exponent $\eta$ is less than 1, implying that spin autocorrelation function $q$($t$) is well approximated by a power law rather than a stretched-exponential or exponential ($\eta >$ 1)~\cite{MacLaughlin_2004}. A value of $\eta$ = 1/2 is also predicted by the mean-field model of a disordered Kondo alloy at a QCP~\cite{PhysRevB.60.4702,PhysRevB.99.224424}.

Similar time field scaling has also been observed in chemically substituted systems near a QCP like UCu$_{5-x}$Pd$_x$, CePtSi$_{1-x}$Ge$_x$, CePd$_{0.15}$Rh$_{0.85}$, or in stoichiometric NFL system CeRhBi \cite{MacLaughlin_2004,PhysRevB.78.014412,doi:10.7566/JPSJ.87.064708} as well as in the spin glass system AgMn (0.5 at \%) above $T_\mathrm{g}$ \cite{PhysRevB.64.054403}. The observed value of $\eta$ = 0.22 for CeRh$_4$Al$_{15}$ is less than the values observed for UCu$_{3.5}$Pd$_{1.5}$, CeRhBi ($\eta$ = 0.7, 0.8) and CePtSi$_{1-x}$Ge$_x$ ($\eta$ = 1.6 for $x$ = 0 and 0.1), but comparable to the value observed for UCu$_4$Pd ($\eta$ = 0.35).


\begin{figure*}
		\includegraphics[width=\textwidth, keepaspectratio]{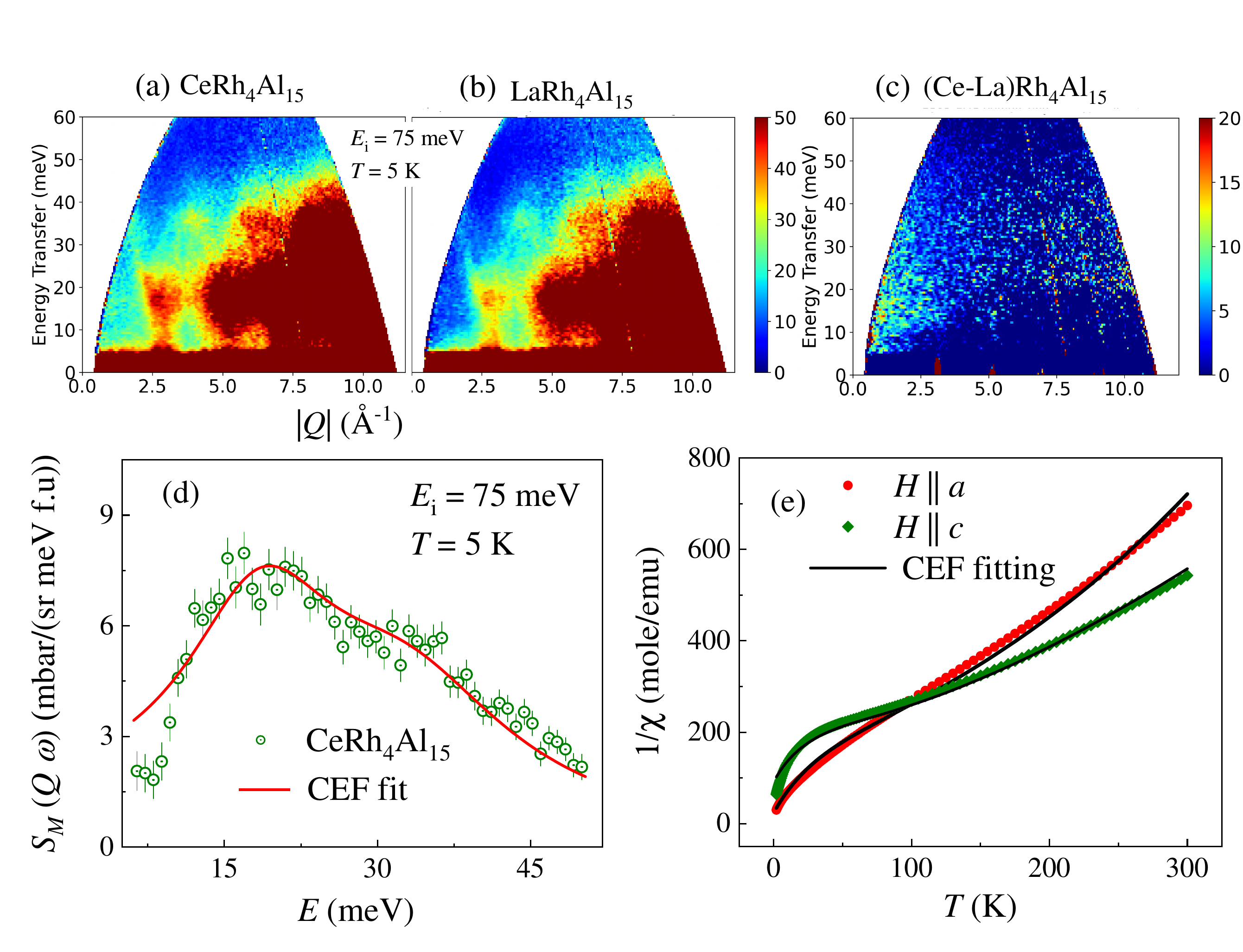}
		\caption{Powder averaged neutron scattering intensity plots measured with incident energy $E_i$=75~meV at temperature $T$ = 5~K for (a) CeRh$_4$Al$_{15}$ (b) LaRh$_4$Al$_{15}$, and (c) The estimated magnetic scattering intensity by subtracting the phonon contribution, CeRh$_4$Al$_{15}$ - LaRh$_4$Al$_{15}$. (d) Magnetic INS response of CeRh$_4$Al$_{15}$ at 5~K after subtracting the phonon contribution from LaRh$_4$Al$_{15}$. The solid line shows the fit based on the CEF model, see text. (e) Single crystal inverse susceptibility versus temperature for the applied field along $a$ and $c$-direction. The solid black lines show the fit based on the CEF model, see text.}
		
		\label{HEINS}
\end{figure*}

In order to investigate the nature of $4f$-electron and single ion CEF anisotropy, we have performed high energy INS measurements on the polycrystalline sample of CeRh$_{4}$Al$_{15}$ ( Fig.~\ref{HEINS}(a). We also measured LaRh$_{4}$Al$_{15}$ to estimate the phonon scattering ( Fig.~\ref{HEINS}(b)). The magnetic scattering of the 75~meV data was estimated using a direct subtraction of the scattering of the non-magnetic reference LaRh$_{4}$Al$_{15}$ from the Ce-data as $S$($Q$,$\omega$)$_{\text{M}}$ = $S$($Q$,$\omega$)$_{\text{Ce}}$ -- $\alpha$ $\times$ $S$($Q$,$\omega$)$_{\text{La}}$. Here $\alpha$ (= 0.87) is the scaling factor obtained from the ratio of the total scattering cross-section of the Ce to La compounds. In this procedure, we found that the phonon modes were still present in the magnetic scattering at high-$Q$. Hence we used $\alpha$ = 1.1  to estimate the magnetic scattering, which resulted in a better estimation of the magnetic scattering. The estimated magnetic scattering is presented in Fig.~\ref{HEINS}(c-d). We also estimated the magnetic scattering using another method ~\cite{PhysRevB.50.9882}, i.e., $S$($Q$,$\omega$)$_{\text{M}}$ = $S$($Q$,$\omega$,low-$Q$)$_{\text{Ce}}$ -- $S$($Q$,$\omega$, high-$Q$)$_{\text{Ce}}$/[$S$($Q$,$\omega$, high-$Q$)$_{\text{La}}$/S($Q$,$\omega$, low-$Q$)$_{\text{La}}$]. We found very similar magnetic scattering (data not shown) as in Fig.~\ref{HEINS}(d). Figure~\ref{HEINS}(d) shows two well defined magnetic excitations centred near 19 and 33~meV, which we attribute due to the CEF excitations of the ground state multiplet  $J=5/2$ of Ce$^{3+}$ splitting into three CEF doublets. The linewidth of the observed CEF excitations is higher than the instrument resolution ($\Delta$$E$=3.8 meV, 3.0 meV at 19 meV and 33 meV energy transfer, respectively). We do not have a clear explanation for this at present, but there are two possibilities; (i) The value of the high-temperature Kondo temperature $T_{\text{K}}^{\text{high}}$ is high ~85~K, which is reflected in the broader linewidth. (ii) As we have some unoccupied Al sites, which will also result in a distribution of the CEF potential and hence a broader linewidth. 

Now we present the analysis of INS data based on the CEF model. The symmetry for the Ce site is approximated by tetragonal point symmetry ($m 2 m$, C$_{4v}$), which results in two CEF excitations in the paramagnetic state. The CEF Hamiltonian for the tetragonal point symmetry ($C_{4v}$) of the Ce$^{3+}$ ions is given by

\begin{equation}\label{H-CEF}
 H_{\rm CEF} = B_{2}^{0}O_{2}^{0} +B_{4}^{0}O_{4}^{0} + B_{4}^{4}O_{4}^{4}
\end{equation}

\noindent where $B_{n}^{m}$ are CEF parameters and $O_{n}^{m}$ are the Stevens operators~\cite{Stevens_1952}. $B_{n}^{m}$ parameters need to be estimated by fitting the experimental data, such as single crystal susceptibility and/or INS data. For the analysis of INS data, we use a Lorentzian line shape for the inelastic excitations.

The $H_{\text{CEF}}$ causes the six-fold degenerate Ce$^{3+}$ ($4f^1, J = 5/2$) state to split into three Kramers doublets. In order to obtain a set of CEF parameters that consistently fit the INS data and single crystal susceptibility, we performed a simultaneous fit of INS at 5~K and the single crystal susceptibility data using the Mantid software~\cite{ARNOLD2014156}. Fits to the INS data at 5~K, and $\chi$($T$) from 5 - 300~K are shown by the solid curve in Fig.~\ref{HEINS}(d) and Fig.~\ref{HEINS}(e), respectively. The CEF parameters obtained from the simultaneous fit are (in meV) $B_2^0$ = -0.6725(5), $B_4^0$ = 0.08515(8), and $B_4^4$ = -0.2576(4). The analysis gives the first excited doublet ($\Delta_1$) at 18.55~meV and the second excited doublet ($\Delta_2$) at 32.67~meV (the fitted values are $\lambda$ = -23(1) and -87(2) and $\chi_0$ = -0.0010(3) emu/mol and -0.0005(1) emu/mol for $H$ along $a$ and $c$ respectively).

\section{Discussion}
Quantum Griffiths phases are generally detected in inhomogeneous systems near a QCP driven by chemical substitution and are responsible for FL breakdown. More surprisingly, here, in stoichiometric CeRh$_{4}$Al$_{15}$, the low temperature NFL behaviors are observed to be favoured by quantum Griffiths scenario, which is described as the power law behavior of (i) the electrical resistivity that increases from a residual value as $\rho \sim T^{\epsilon}$ with (1 $\leq \epsilon<2$); (ii) the Sommerfield coefficient $\gamma=C/T \sim T^{\alpha_{\text{C}}}$; (iii) the dc magnetic susceptibility $\chi \sim T^{\alpha_{\chi}}$; and (iv) the isotherm magnetization, $M \sim H^{\alpha}$. We observe that there is a disparity between the estimated values of $\alpha_{\chi}$, $\alpha_{\text{C}}$, and $\alpha$ inferred from the fits to magnetic susceptibility, heat capacity, and magnetization data, respectively. Similar discrepancies have also been observed by Castro Neto~\cite{PhysRevLett.81.5620}, which were attributed to magnetocrystalline anisotropy and the preferred crystalline orientation in single crystalline samples. Moreover, the scaling analysis of the magnetic field dependent specific heat yields an exponent $\beta$ greater than 1 which rules out a single-ion effect and may be taken as further evidence of Griffiths phase to be a possible origin for the NFL behavior.

Now, we propose that the quantum Griffiths phase in CeRh$_{4}$Al$_{15}$ emerges from the defects in the crystal structure as the Al sites are not 100\% occupied, which locally modify the exchange interactions in this undoped compound and also disorder in CEF potential. The latter has been observed/confirmed through broad CEF excitations. We anticipate that such defects could result in the formation of magnetic clusters in proximity to the QCP leading to NFL behavior, such as those resulting from impurities or bond disorders induced quantum Griffiths phase proposed by Castro Neto $et al$.~\cite{PhysRevLett.81.3531,PhysRevLett.81.5620}.

\section{Conclusions}

In conclusion, we have reported a comprehensive study of electrical transport, magnetic susceptibility, and heat capacity measurements on a single crystal sample of CeRh$_{4}$Al$_{15}$ together with $\mu$SR and neutron scattering experiments on the polycrystalline sample. A significant deviation of the physical properties from a FL behavior, such as $\rho \sim T^{\epsilon}(1 \leq \epsilon<2)$, $C / T \sim T^{\alpha_{\text{C}}}$, $\chi \sim T^{\alpha_{\chi}}$, and $M \sim H^{\alpha}$ with $\alpha_{\text{C}} \sim \alpha_{\chi} \sim \alpha$ $< 1$, is observed, which has been attributed to the NFL behavior in proximity to the QCP. The observed value of the exponents $\alpha<1$ are consistent with the power-law Griffiths singularity proposed by Castro Neto $et~ al$.~\cite{PhysRevLett.81.3531,PhysRevLett.81.5620}. Moreover, the scaling analysis of the magnetic field dependent specific heat yields an exponent $\beta$ ($>1$) that may be taken as further evidence of Griffiths phase as a possible origin for the NFL behavior.

The temperature dependence of the ZF-$\mu$SR dynamic relaxation rate $\lambda$ exhibits a thermal activation-like characteristic [$T$log $\lambda$ $\sim$ $T$] over the entire measured temperature range, indicating the presence of low energy spin fluctuations in CeRh$_4$Al$_{15}$. The LF-$\mu$SR data at 100~mK exhibit a time-field scaling with the exponent $\eta = 0.22$(1), which suggests that the spin-spin autocorrelation function has a power-law behavior. INS study shows two broad CEF excitations. The simultaneous analysis of INS and single crystal susceptibility based on the CEF model explains the observed cross-over behavior of the temperature dependence of single crystal susceptibility between $a$- and $c$-axes.

The XRD analysis further confirms that the structural disorder is due to partial occupancy on some of the Al sites, which could affect the physical properties and hence results in the observed behavior. We, thus, propose that Griffiths phase crucially controls the low-temperature spin dynamics and is responsible for NFL behavior respectively. Our study provides evidence of the presence of such clusters inside a paramagnetic environment even in the undoped compound.

\begin{acknowledgments}
We gratefully acknowledge the ISIS facility for the beam time on MERLIN (RB1920702)~\cite{MERLIN}, EMU (RB2010778) ~\cite{EMU}. DTA would like to thank the Royal Society of London for International Exchange funding between the UK and Japan, Newton Advanced Fellowship funding between UK and China and EPSRC UK for the funding (Grant No. EP/W00562X/1). RT thanks the Indian Nanomission for a post-doctoral fellowship. AMS thanks the SA-NRF and the URC/FRC of UJ for financial assistance.

\end{acknowledgments}

\bibliography{bibliography}

\end{document}